\documentclass[aps,prb,twocolumn,superscriptaddress,longbibliography]{revtex4-1}
\usepackage{amssymb, amsmath, bm}
\usepackage{graphicx,onlyamsmath}

\usepackage{natbib}
\usepackage{xcolor}
\usepackage[normalem]{ulem}  
\usepackage{makecell} 

\begin{document}

\title{Disorder-induced phase transitions in double HgTe quantum wells}

\author{S. S.~Krishtopenko}
\thanks{These authors contributed equally to this work}
\affiliation{Laboratoire Charles Coulomb (L2C), UMR 5221 CNRS-Universit\'{e} de Montpellier, F-34095 Montpellier, France}

\author{A. V.~Ikonnikov}
\thanks{These authors contributed equally to this work}
\affiliation{Physics Department, M.V. Lomonosov Moscow State University, Moscow, 119991, Russia}

\author{B.~Jouault}
\affiliation{Laboratoire Charles Coulomb (L2C), UMR 5221 CNRS-Universit\'{e} de Montpellier, F-34095 Montpellier, France}

\author{F.~Teppe}
\email[]{frederic.teppe@umontpellier.fr}
\affiliation{Laboratoire Charles Coulomb (L2C), UMR 5221 CNRS-Universit\'{e} de Montpellier, F-34095 Montpellier, France}

\date{\today}

\begin{abstract}
By using the self-consistent Born approximation, we investigate topological phase transitions in double HgTe quantum wells (QWs) induced by the short-range impurities. Following the evolution of the density-of-states and the spectral function, we demonstrate multiple closings and openings of the band-gap with the increase of the disorder strength due to the mutual inversions between the first and second electron-like and hole-like subbands. We show that starting from a band insulator in the clean limit, under the influence of disorder, the double HgTe QW undergoes a transition, first into a semimetal state similar to ``bilayer graphene'', and then into a high-order topological insulator state with a double band inversion. We find out that all disorder-induced transitions can be fully characterized by introducing a non-Hermitian quasiparticle Hamiltonian encoding the band structure renormalization and quasiparticle decay.
\end{abstract}

\keywords{}
\maketitle

\section{\label{Sec:Int} Introduction}
The band structure of HgTe quantum wells (QWs) in the vicinity of the $\Gamma$ point of the Brillouin zone is represented by the Dirac fermions with additional terms quadratic in quasimomentum~\cite{q1}. The inherent property of these fermions is that their band-gap can be varied from positive to negative values and vice versa by changing the QW width~\cite{q2,q3,q4}, temperature~\cite{q5,q6,q7,q8} or hydrostatic pressure~\cite{q9}. The negative band-gap values in HgTe QWs correspond to the so-called \emph{inverted} band structure, in which the lowest electron-like level \emph{E}1 lies below the top hole-like level \emph{H}1, resulting in the emergence of the quantum spin Hall insulator (QSHI) state~\cite{q10,q11}. This topologically non-trivial state is characterized by a bulk band-gap and a pair of helical gapless states at the edges that leads to quantized values of the conductance. This edge conductance quantization without an applied magnetic field represents the ultimate experimental confirmation of the non-trivial topology of a 2D system~\cite{q11,q12,q13}.

Unexpectedly, Li~\emph{et~al.}~\cite{q14} discovered by numerical simulations that initially clean HgTe QWs with trivial band ordering can also feature a quantized conductance value in the presence of disorder. This disordered state with quantized conductance is referred to as a \emph{topological Anderson insulator} (TAI). Later, Groth~\emph{et~al.}~\cite{q15} explained the TAI mechanism as a disorder-induced band inversion arising due to the presence of quadratic terms in the Dirac-like low-energy Hamiltonian of HgTe QWs~\cite{q1}. In this sense, the phenomenology of TAI should be very similar to that of the QSHI. Interestingly, the discovered transition correspond to weak-disorder topological transition, and thus can be treated within the self-consistent Born approximation (SCBA)~\cite{q15,q16,q17,q17b,q18}. By now, disorder-induced phase transitions have also been studied in 3D topological insulators~\cite{TAI3Da,TAI3Db,TAI3Dc}, various topological semimetals~\cite{TAI3Dd,TAI3De,TAI3Df,TAI3Dg,TAI3Dh}, amorphous solids~\cite{TAIAm}, chiral \emph{p}-wave superconductors~\cite{TAIAm1} and even in mechanical systems~\cite{TAIAm2}.

A distinctive feature of all these works~\cite{TAI3Da,TAI3Db,TAI3Dc,TAI3Dd,TAI3De,TAI3Df,TAI3Dg,TAI3Dh,TAIAm} is that the theoretical investigations were performed within the framework of the so-called two-band approximation, which allows describing the disorder-induced evolution of the conduction and valence band edges only. In this case, questions naturally arise: (i) how the disorder affects the edges of the higher-lying conduction and valence bands beyond the two-band model; and (ii) whether the disorder may cause a double-band inversion between second electron- and hole-like bands. These double-band-inverted states are now of interest in the context of higher-order topology~\cite{q19,q20,q21}.

In this work, considering double HgTe QWs as a seminal example, we study disorder-induced phase transitions in a 2D system, whose phase diagram contains trivial, single-band-inverted and double-band-inverted states~\cite{q22} (see Fig.~\ref{Fig:1}). Different types of these states were previously probed in realistic samples by magnetotransport~\cite{Tr1,Tr2,Tr3,Tr4,Tr5,Tr6,Tr7} and mid-infrared Landau level spectroscopy~\cite{LL1,LL1b,LL2,LL2b,LL3,LL4}. By using the SCBA and four-band 2D low-energy Hamiltonian, describing topological phase transitions in double HgTe QWs~\cite{q22,q23}, we directly calculate the DOS and spectral function visualizing quasiparticle picture at the $\Gamma$ point of the Brillouin zone. By following the DOS evolution with increasing of the disorder strength, we unambiguously demonstrate multiple topological phase transitions caused by the mutual inversion of both the first and second electron-like and hole-like subbands. We find out that all disorder-induced transitions are completely characterized by introducing a non-Hermitian quasiparticle Hamiltonian encoding the band structure renormalization and quasiparticle decay. We consider the electrostatic potential of randomly distributed short-range impurities as a source of the disorder, which makes it possible to reduce the system of self-consistent integral equations to an algebraic one.

\section{\label{Sec:Diagram} Phase diagram in the clean limit}
\begin{figure}
\includegraphics [width=1.0\columnwidth, keepaspectratio] {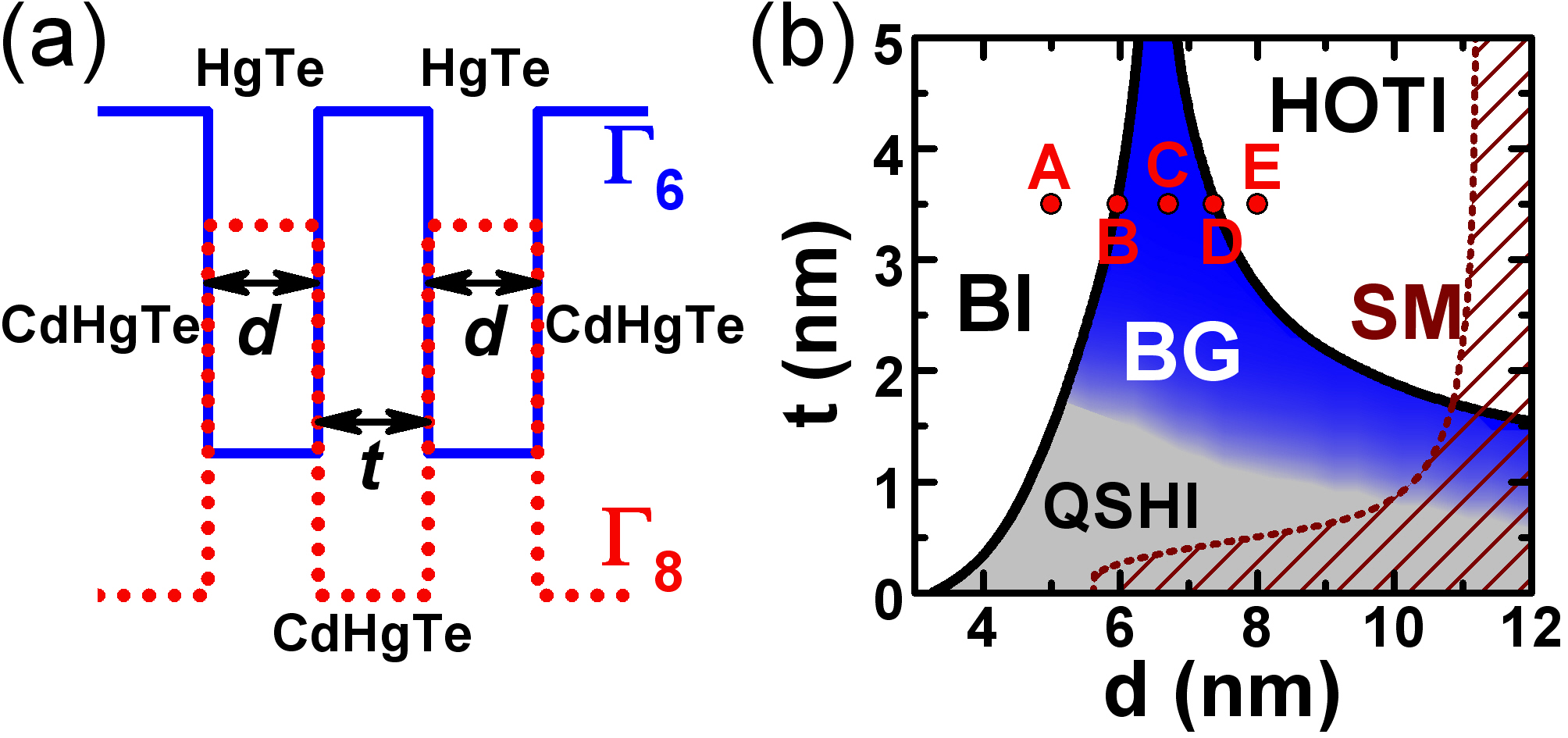} 
\caption{\label{Fig:1} (a) Schematic representation of \emph{symmetric} double HgTe~QW. Here, $d$ is the thickness of HgTe layers and $t$ is the middle CdHgTe barrier thickness. The double QW is assumed to be grown on (001) CdTe buffer with 30\% of Hg content in all the barriers~\cite{q22}. (b) The phase diagrams for different $d$ and $t$ obtained by numerical calculations based on eight-band \textbf{k$\cdot$p} Hamiltonian~\cite{q9}. The left-hand and right-hand solid curves correspond to the crossing between \emph{E}1--\emph{H}1 subbands and \emph{E}2--\emph{H}2 subbands, respectively. These curves divide the plane into three parts with trivial band ordering (left-hand white region), single band inversion (grey and blue regions) and double-band inversion (right-hand white region). The striped region defines a semimetal (SM) phase with vanishing \emph{indirect} band-gap. The red symbols mark the double QWs considered in this work.}
\end{figure}

Let us first overview the phase diagram of double HgTe QWs in the absence of disorder. In what follows, we consider a symmetrical QW grown on (001) CdTe buffer consisting of two $d$ thick HgTe layers separated by a $t$ thick Cd$_{0.7}$Hg$_{0.3}$Te barrier (see Fig.~\ref{Fig:1}). The phase diagram of double HgTe QWs was obtained on the basis of eight-band \textbf{k$\cdot$p} Hamiltonian~\cite{q9} for the envelope wave functions, which takes into account the interaction between $\Gamma_6$, $\Gamma_8$, and $\Gamma_7$ bands in bulk materials. Note that possible terms resulting from the bulk inversion asymmetry of the unit cell of zinc-blende semiconductors~\cite{q24,q24b,q24c} and the anisotropy of chemical bonds at the QW interfaces~\cite{q25} are neglected in the calculations. These terms are small in HgTe-based heterostructures~\cite{q3,q27,q28,q29,q30,q31}, and in their absence, energy band dispersion is double spin degenerate for all quasimomentum values. The calculation details and material parameters can be found in Ref.~\cite{q22}.

\begin{figure*}
\includegraphics [width=2.05\columnwidth, keepaspectratio] {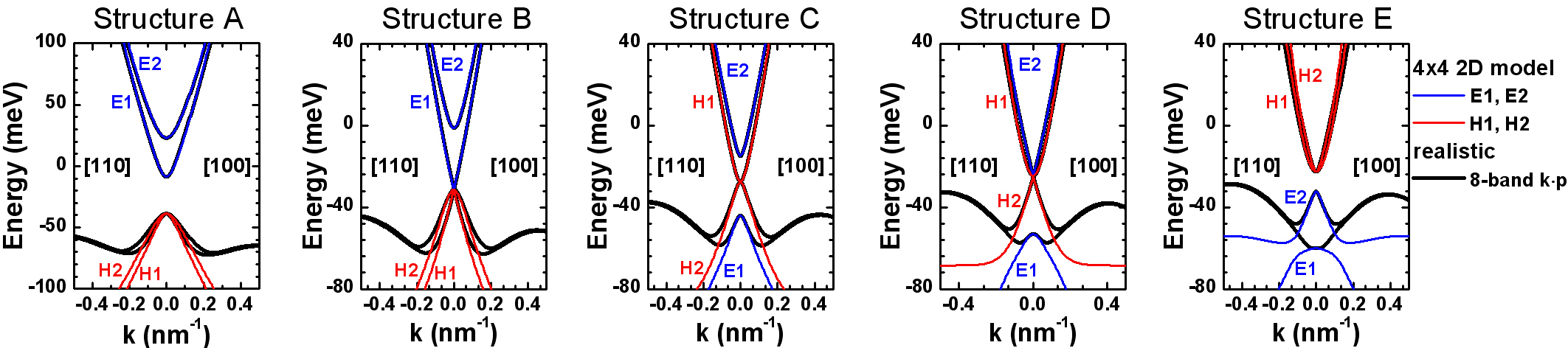} 
\caption{\label{Fig:2} Band structure of double HgTe QWs with $t=3.5$~nm. The values of $d$ are marked by red symbols in Fig.~\ref{Fig:1}. The bold black curves correspond to realistic band structure calculations on the basis of eight-band \textbf{k$\cdot$p} Hamiltonian~\cite{q9}. The thin curves represent the energy-dispersion for electron-like (in blue) and hole-like (in red) subbands calculated by using the \emph{isotropic part} of $H_{2D}(\mathbf{k})$ in Eq.~(\ref{eq:1}). Note that each subband is doubly spin-degenerate. The positive and negative values of $\mathbf{k}$ correspond to the [100] and [110] crystallographic orientations, respectively. The parameters for $H_{4\times4}(\mathbf{k})$ are provided in Table~\ref{Tab:1}.}
\end{figure*}

Unlike single HgTe QWs~\cite{q1}, the low-energy physics of double HgTe QWs is determined by the mutual arrangement of four levels at once -- two electron-like (\emph{E}1, \emph{E}2) and two hole-like (\emph{H}1, \emph{H}2) subbands. Depending on the thickness of the HgTe layers and the width of the tunneling barrier, different subband arrangements at the $\Gamma$ point of the Brillouin zone can be implemented. In the diagram of double HgTe QWs, the left-hand black solid curve representing the crossing between the first electron-like \emph{E}1 and hole-like \emph{H}1 subbands divides the $d$-$t$ plane into a white region, corresponding to band insulator (BI) with trivial band ordering, and a grey region of QSHI with inverted band structure. If the middle barrier is thick enough, in addition to QSHI, the double HgTe QW also hold a specific state with a band structure similar to the one of bilayer graphene (BG) represented by the blue region in Fig.~\ref{Fig:1}(b). The BG state with the \emph{E}2-\emph{H}1-\emph{H}2-\emph{E}2 level sequence (see Fig.~\ref{Fig:2}(c)) was previously discussed in details in Ref.~\cite{q22}.

A further increase of $d$ results in the band crossing between the second electron-like (\emph{E}2) and hole-like (\emph{H}2) subbands, which is represented by the right-hand black solid curve in the diagram. This curve, in its turn, separates the grey and blue regions with single band inversion from the right-hand white region corresponding to the double band inversion, when two electron-like \emph{E}1 and \emph{E}2 levels lie below two hole-like \emph{H}1 and \emph{H}2 subbands (the \emph{H}1-\emph{H}2-\emph{E}2-\emph{E}1 sequence). Recently, such double-inversion insulator state has been identified as a higher-order topological insulator (HOTI)~\cite{q23} with the corner states directly attributed to the cubic symmetry of zinc-blende semiconductors. Finally, at certain $d$ and $t$ values corresponding to the striped region, the so-called semimetal (SM) phase is implemented. The SM state is characterized by a vanishing \emph{indirect} band-gap when the side maxima of the valence subband exceed in energy the conduction subband bottom~\cite{q24}. Thus, by varying the layer thicknesses in double HgTe QWs, one can indeed realize BI, QSHI, BG state, HOTI or SM state. Typical band dispersions of double HgTe QWs arising at different thicknesses of the HgTe layer are shown in Fig.~\ref{Fig:2}.

The multiple band inversions in double HgTe QWs at the $\Gamma$ point of the Brillouin zone can be explicitly described by means of effective four-band 2D low-energy Hamiltonian~\cite{q22}. Within the following sequence of the basis states $|E1{+}\rangle$, $|H1{+}\rangle$, $|H2{-}\rangle$, $|E2{-}\rangle$, $|E1{-}\rangle$, $|H1{-}\rangle$, $|H2{+}\rangle$, $|E2{+}\rangle$, the effective 2D Hamiltonian for the states in the vicinity of the $\Gamma$ point has the form~\cite{q23}:
\begin{equation}
\label{eq:1}
H_{2D}(\mathbf{k})=\begin{pmatrix}
H_{4\times4}(k_x,k_y) & 0 \\ 0 & H_{4\times4}^{*}(-k_x,-k_y)\end{pmatrix},
\end{equation}
where $\mathbf{k}=\left(k_x,k_y\right)$ is quasimomentum in the QW plane and the asterisk denotes complex conjugation. Note that since $H_{2D}(\mathbf{k})$ preserves inversion symmetry, its eigenvalues are double degenerate at all values of $\mathbf{k}$. The diagonal blocks of $H_{2D}(\mathbf{k})$, in turn, are split into isotropic and anisotropic parts:
\begin{equation}
\label{eq:2}
H_{4\times4}(\mathbf{k})=H_{4\times4}^{(i)}(\mathbf{k})+H_{4\times4}^{(a)}(\mathbf{k}).
\end{equation}

The isotropic term $H_{4\times 4}^{(i)}(\mathbf{k})$ preserves the rotational symmetry in the QW plane, therefore it is independent of the orientation of $x$ and $y$ axis. The isotropic part of $H_{4\times4}(\mathbf{k})$ in Eq.~(\ref{eq:2}) is written as~\cite{q22,q22b}:
\begin{equation}
\label{eq:3}
H_{4\times4}^{(i)}(\mathbf{k})=\begin{pmatrix}
\epsilon_{E1} & -A_{1}k_{+} & R_{1}^{(i)}k_{-}^2 & S_{0}k_{-}\\
-A_{1}k_{-} & \epsilon_{H1} & 0 & R_{2}^{(i)}k_{-}^2\\
R_{1}^{(i)}k_{+}^2 & 0 & \epsilon_{H2}  & A_{2}k_{+}\\
S_{0}k_{+} & R_{2}^{(i)}k_{+}^2 & A_{2}k_{-} & \epsilon_{E2} \end{pmatrix},
\end{equation}
where
\begin{eqnarray}
\label{eq:4}
\epsilon_{E1}(k_x,k_y)=C_1+M_1-(D_{1}+B_{1})(k_x^2+k_y^2),~\nonumber\\
\epsilon_{H1}(k_x,k_y)=C_1-M_1-(D_{1}-B_{1})(k_x^2+k_y^2),~\nonumber\\
\epsilon_{E2}(k_x,k_y)=C_2+M_2-(D_{2}+B_{2})(k_x^2+k_y^2),~\nonumber\\
\epsilon_{H2}(k_x,k_y)=C_2-M_2-(D_{2}-B_{2})(k_x^2+k_y^2).~
\end{eqnarray}
Here, $k_{\pm}=k_x+ik_y$, $k_x$ and $k_y$ are the momentum components in the QW plane, and $C_{1,2}$, $M_{1,2}$, $A_{1,2}$, $B_{1,2}$, $D_{1,2}$, $S_0$ and $R_{1,2}^{(i)}$ are \emph{isotropic} structure parameters being defined by the QW geometry, the growth orientation and the layer materials. As can be seen from their physical meaning, the parameters $S_0$ and $R_{1,2}^{(i)}$, which describe the coupling between the HgTe layers, vanish if $t\rightarrow\infty$.

On the contrary, the form of $H_{4\times4}^{(a)}(\mathbf{k})$ in Eq.~(\ref{eq:2}), resulting from the cubic symmetry of zinc-blende semiconductors, strongly depends on the QW growth orientation but also on the orientation of $x$ and $y$ axes. Assuming the $x$ and $y$ axes are oriented along the crystallographic directions (100) and (010), respectively, $H_{4\times4}^{(a)}(\mathbf{k})$ in ($001$)-oriented QWs has the form~\cite{q23}
\begin{equation}
\label{eq:4ad2}
{H}_{4\times4}^{(a)}(\mathbf{k})=-\begin{pmatrix}
0 & 0 & R_{1}^{(a)}{k}_{+}^2 & 0\\
0 & 0 & 0 & R_{2}^{(a)}{k}_{+}^2\\
R_{1}^{(a)}{k}_{-}^2 & 0 & 0  & 0\\
0 & R_{2}^{(a)}{k}_{-}^2 & 0 & 0 \end{pmatrix}
\end{equation}
where $R_{1,2}^{(a)}$ are \emph{cubic} structure parameters, which depend on the QW geometry and composition. It is important to note that since $\left|R_{1,2}^{(a)}\right|{\ll}\left|R_{1,2}^{(i)}\right|$,  $H_{4\times4}^{(a)}(\mathbf{k})$ does not have a significant effect on the qualitative picture of 2D bulk and 1D edge states in double HgTe QWs. However, taking $H_{4\times4}^{(a)}(\mathbf{k})$ into account is critical for the mass kink of the gapped Dirac edge states as function of the edge orientation under double band inversion (see right-hand white region in Fig.~\ref{Fig:1}(b)). This results in the appearance of a localized 0D corner state at the meeting edges of certain crystallographic orientations~\cite{q23}.

Figure~\ref{Fig:2} compares the band structure calculations based on the realistic eight-band \textbf{k$\cdot$p} Hamiltonian~\cite{q9} and $H_{2D}(\mathbf{k})$ in Eq.~(\ref{eq:1}) for double HgTe QWs with $t=3.5$~nm at several values of $d$ (marked by the red symbols in the phase diagram of Fig.~\ref{Fig:1}). There is a good agreement between the results of both models for small quasimomentum values. The structural parameters used in the calculations are summarized in Tab.~\ref{Tab:1}.  Note that since the SM phase represented by the striped areas in the diagram of Fig.~\ref{Fig:1}(b) is formed by an overlap at non-zero quasimomentum of the valence and conduction subbands, it cannot be described within the low-energy Hamiltonian valid for the states at the small values of $k_x$ and $k_y$.

The most important quantities in $H_{2D}(k_x,k_y)$ are the two mass parameters $M_1$ and $M_2$ describing the band inversion between \emph{E}1--\emph{H}1 subbands and \emph{E}2--\emph{H}2 subbands, respectively. Changing the sign of one of these two parameters leads to a topological phase transition. Indeed, the parity of subbands at the $\Gamma$ point involved into $H_{2D}(\mathbf{k})$ are define as~\cite{q22}
\begin{eqnarray}
\label{eq:4ad3}
\mathcal{P}|E1{\pm}\rangle=-|E1{\pm}\rangle,~~~~~\mathcal{P}|E2{\pm}\rangle=|E2{\pm}\rangle,~~\notag\\
\mathcal{P}|H1{\pm}\rangle=|H1{\pm}\rangle,~~~~~~\mathcal{P}|H2{\pm}\rangle=-|H2{\pm}\rangle,
\end{eqnarray}
where $\mathcal{P}$ is the inversion symmetry operator.
It is easy to see that when the sign of $M_1$ or $M_2$ changes, the parity of the valence (occupied) subbands changes as well, which, in accordance with the Fu-Kane criterion~\cite{q36}, leads to a change in the $\mathbb{Z}_2$ invariant. Let us emphasize that the effective four-band Hamiltonian also allows for extended description of single HgTe QWs~\cite{q22b} than within the Bernevig-Hughes-Zhang (BHZ) model~\cite{q1}. Therefore, the parity of the valence (occupied) subbands in double HgTe QWs can be compared to the case of single HgTe QWs, for which the phase diagram is well-known.

Note that the calculation of the $\mathbb{Z}_2$ invariant requires the knowledge of the parity of the valence subband eigenstates at all time-reversal invariant momenta $\mathbf{k_0}$ in the Brillouin zone:
\begin{equation}
\label{eq:4ad4}
(-1)^{\mathbb{Z}_2}=\prod\limits_{\mathbf{k_0}}\prod\limits_{n:~\mathrm{valence}}
\pi_{n}\left(\mathbf{k_0}\right),
\end{equation}
where $\pi_{n}\left(\mathbf{k_0}\right)\in\{\pm{1}\}$ is the parity eigenvalue of the $n$-th Kramers pair at $\mathbf{k}=\mathbf{k_0}$. The fact that in HgTe-based QWs the subband inversion occurs only at the $\Gamma$ point of the Brillouin zone indicates that the product of the parities of the valence subbands at other time-reversal invariant momenta does not change. Therefore, in what follows we focus only on the parity change at the $\Gamma$ point.

As clear from the comparison with single HgTe QWs, positive values of $M_1$ and $M_2$ correspond to trivial BI with $\mathbb{Z}_2=0$. In this case, the product of the parities of valence \emph{H}1 and \emph{H}2 subbands is equal to $-1$. The latter means that the parity product of the \emph{remote valence subbands}, which are not included in $H_{2D}(\mathbf{k})$, equals to $-1$. This allows reducing Eq.~(\ref{eq:4ad4}) to
\begin{equation}
\label{eq:4ad5}
-(-1)^{\mathbb{Z}_2}=\prod\limits_{n:~\mathrm{valence~of~}{H_{2D}(\mathbf{k})}}
\pi_{n}\left(0\right).
\end{equation}
The QSHI and BG states in double HgTe QWs at $M_1<0$ and $M_2>0$ has the same product of valence subband parities of $+1$ as the QSHI state in single HgTe QWs~\cite{q22b}. Thus, the QSHI and BG states are both characterized by $\mathbb{Z}_2=1$, and the only difference between these states is defined by the gap between the \emph{H}1 and \emph{H}2 subbands~\cite{q22}. Finally, the product of the parities of the valence subbands for negative values of $M_1$ and $M_2$ is equal to $-1$, which means that $\mathbb{Z}_2=0$, as it is for the BI state. In order to identify a HOTI state in the insulator with double band inversion, one should take into account the crystal symmetry effect described by ${H}_{4\times 4}^{(a)}(\mathbf{k})$. The latter results in 0D states arising at the corners of the meeting edges, whose bisectors coincide with the mirror symmetry planes of (001) double HgTe QWs~\cite{q23}.

\begin{table*}
\caption{\label{Tab:1} Structure parameters involved in the \emph{isotropic part} of $H_{4\times4}(\mathbf{k})$ in Eq.~(\ref{eq:2}) for double HgTe/CdHgTe QWs, whose layer thicknesses are represented by red symbols in Fig.~\ref{Fig:1}. We also note that $C_2=C_1-M_1+M_2$. The structure parameter for an arbitrary thickness of the HgTe layers in the range from $5.0$~nm to $8.00$~nm can be obtained by interpolation between the values presented in the table. The structural parameters of the \emph{anisotropic part} of $H_{4\times4}(\mathbf{k})$ can be found in Ref.~\cite{q23}.}
\begin{ruledtabular}
\begin{tabular}{c|c|c|c|c|c|c|c|c}
Structure & $d$ (nm) & $t$ (nm) & $C_1$ (meV) & $M_1$ (meV) & $M_2$ (meV) & $B_1$ (meV$\cdot$nm$^2$) & $B_2$ (meV$\cdot$nm$^2$) & $D_1$ (meV$\cdot$nm$^2$) \\
\hline
A & 5.00 & 3.50 & -23.73  & 15.04 & 30.91 & -494.80 & -397.54 & -422.00 \\
B & 5.97 & 3.50 & -31.37 & 0.00 & 15.00 & -669.41 & -537.84 & -600.85  \\
C & 6.70 & 3.50 & -35.73 & -8.32 & 6.25 & -851.05 & -615.76 & -786.85  \\
D & 7.37 & 3.50 & -38.88 & -14.20 & 0.00 & -1071.75 & -694.25 & -1012.46  \\
E & 8.00 & 3.50 & -41.34 & -18.74 & -4.82 & -1431.86 & -777.63 & -1378.16
\end{tabular}
~\\~
\begin{tabular}{c|c|c|c|c|c|c}
Structure & $D_2$ (meV$\cdot$nm$^2$) & $A_1$ (meV$\cdot$nm) & $A_2$ (meV$\cdot$nm) & $R_{1}^{(i)}$ (meV$\cdot$nm$^2$) & $R_{2}^{(i)}$ (meV$\cdot$nm$^2$) & $S_0$ (meV$\cdot$nm) \\
\hline
A & -321.67 & 405.08 & 391.19 & 67.50 & -92.96 & -2.94 \\
B & -465.37 & 387.65 & 372.25 & 90.64 & -108.30 & 0.05  \\
C & -546.76 & 374.07 & 358.11 & 116.53 & -121.70 & 2.43  \\
D & -629.15 & 361.40 & -345.31 & -405.09 & -135.09 & 4.73  \\
E & -716.99 & 348.99 & -333.11 & -501.63 & -148.81 & 7.04
\end{tabular}
\end{ruledtabular}
\end{table*}

\section{\label{Sec:SCBA} SCBA for double HgTe QWs}
So far, we have discussed the ``clean'' case of double HgTe QWs. In this section, we focus on the disorder effect on the band ordering in double HgTe QWs, caused by randomly distributed impurities in the QW plane. Since we are interested in the topological phase transitions induced by weak disorder, we will treat them within the self-consistent Born approximation~\cite{q15,q16,q17,q18}. Further, we represent the calculations for the upper block of $H_{2D}(\mathbf{k})$ in Eq.~(\ref{eq:1}), while the calculations for the lower block are performed in the same way.

Let us consider Green's function defined by
\begin{equation}
\label{eq:5}
\hat{G}(\mathbf{k},\varepsilon)=\langle\dfrac{1}{\varepsilon-\mathcal{H}}\rangle=
\left[\varepsilon-H_{4\times4}(\mathbf{k})-\hat{\Sigma}(\mathbf{k},\varepsilon)\right]^{-1},
\end{equation}
with
\begin{equation}
\label{eq:6}
\mathcal{H}=H_{4\times4}(\mathbf{k})+V_{imp}(\textbf{r}),
\end{equation}
where $\langle...\rangle$ denotes average over all disorder configurations, and $\hat{\Sigma}(\mathbf{k},\varepsilon)$ is the self-energy matrix. In Eq.~(\ref{eq:6}), we have also introduced a disorder potential $V_{imp}(\textbf{r})$, consisting of randomly distributed individual impurities:
\begin{eqnarray}
\label{eq:7}
V_{imp}(\textbf{r})=\sum_{j}v(\textbf{r}-\textbf{R}_j),~v(\textbf{r})=\int\dfrac{d^2\textbf{q}}{(2\pi)^2}\tilde{v}(\textbf{q})e^{i\textbf{q}\cdot\textbf{r}},~~~~~
\end{eqnarray}
where $R_j$ denotes position of impurities and $v(\textbf{r})$ is the potential of an individual impurity, which is assumed to be isotropic, i.e., $\tilde{v}(\textbf{q})=\tilde{v}(q)$ with $|\textbf{q}|=q$.

As noted in the previous section, anisotropic structural parameters $R_{1}^{(a)}$ and $R_{2}^{(a)}$ are significantly smaller than isotropic $R_{1}^{(i)}$ and $R_{2}^{(i)}$. Therefore, since $H_{4\times4}^{(i)}(\mathbf{k})$ and $H_{4\times4}^{(a)}(\mathbf{k})$ have the same order in quasimomentum, we further neglect the small contribution of $H_{4\times4}^{(a)}(\mathbf{k})$ when describing the disorder effect on the band ordering in double HgTe QWs. This allows one to significantly simplify mathematical calculations. Indeed, in the absence of $H_{4\times4}^{(a)}(\mathbf{k})$, $H_{4\times4}(\mathbf{k})$ possesses full rotational symmetry, and its wave-function can be presented in the form:
\begin{equation}
\label{eq:8}
\Psi_{4\times4}(\mathbf{k})=U(\theta_{\mathbf{k}})^{-1}\Psi_{4\times4}(k),
\end{equation}
where $k=|\textbf{k}|$, $k_x=k\cos\theta_{\mathbf{k}}$, $k_y=k\sin\theta_{\mathbf{k}}$,
\begin{equation}
\label{eq:9}
U(\theta)=\begin{pmatrix}
1 & 0 & 0 & 0 \\
0 & e^{i\theta} & 0 & 0 \\
0 & 0 & e^{-2i\theta} & 0 \\
0 & 0 & 0 & e^{-i\theta} \end{pmatrix}.
\end{equation}
Therefore, the Green's function in Eq.~(\ref{eq:5}) can be presented in the form
\begin{equation}
\label{eq:10}
\hat{G}(\mathbf{k},\varepsilon)=U(\theta_{\mathbf{k}})\hat{G}(k,\varepsilon)U(\theta_{\mathbf{k}})^{-1},
\end{equation}
with
\begin{equation}
\label{eq:11}
\hat{G}(k,\varepsilon)=\left[\varepsilon-\tilde{H}_{4\times4}(k)-\hat{\Sigma}(k,\varepsilon)\right]^{-1},
\end{equation}
which depends only on $k$. This shows that $\hat{G}(\mathbf{k},\varepsilon)$ depends on the angle via the terms of $U(\theta_{\mathbf{k}})$. We note that $\tilde{H}_{4\times4}(k)$ differs from $H_{4\times4}(\mathbf{k})$ by
\begin{equation}
\label{eq:12}
\tilde{H}_{4\times4}(k)=U(\theta_{\mathbf{k}})H_{4\times4}(\mathbf{k})U(\theta_{\mathbf{k}})^{-1}.
\end{equation}

Within the SCBA, the self-energy matrix in Eq.~(\ref{eq:5}) has a form
\begin{equation}
\label{eq:13}
\hat{\Sigma}(\textbf{k},\varepsilon)=n_{i}\int\dfrac{d^2\textbf{k}^\prime}{(2\pi)^2}
\tilde{v}(\textbf{k}-\textbf{k}^\prime)\hat{G}(\mathbf{k}^\prime,\varepsilon)\tilde{v}(\textbf{k}^\prime-\textbf{k}),
\end{equation}
where $n_{i}$ is the concentration of impurities. Similar to Eq.~(\ref{eq:8}), the self-energy matrix can be represented as
\begin{equation}
\label{eq:14}
\hat{\Sigma}(\mathbf{k},\varepsilon)=U(\theta_{\mathbf{k}})\hat{\Sigma}(k,\varepsilon)U(\theta_{\mathbf{k}})^{-1},
\end{equation}
with a matrix $\hat{\Sigma}(k,\varepsilon)$ written as
\begin{eqnarray}
\label{eq:15}
\hat{\Sigma}(k,\varepsilon)=n_{i}\int\limits_0^{K_c}\dfrac{k^\prime dk^\prime}{2\pi}~~~~~~~~~~~~~~~~~~~~~~~~~~~~~~~~~~~\notag\\
\times\begin{pmatrix}
V_0^2G_{11}^\prime & V_{-1}^2G_{12}^\prime & V_{+2}^2G_{13}^\prime & V_{+1}^2G_{14}^\prime \\[2pt]
V_{+1}^2G_{21}^\prime & V_{0}^2G_{22}^\prime & V_{+3}^2G_{23}^\prime & V_{+2}^2G_{24}^\prime \\[2pt]
V_{-2}^2G_{31}^\prime & V_{-3}^2G_{32}^\prime & V_{0}^2G_{33}^\prime & V_{-1}^2G_{34}^\prime \\[2pt]
V_{-1}^2G_{41}^\prime & V_{-2}^2G_{42}^\prime & V_{+1}^2G_{43}^\prime & V_{0}^2G_{44}^\prime
\end{pmatrix}.~~
\end{eqnarray}
Here, $G_{ij}^\prime{\equiv}G_{ij}(k^\prime,\varepsilon)$ are the component of the averaged Green's function in Eq.~(\ref{eq:11}), and $V_n^{2}{\equiv}V_n(k,k^\prime)^2$ are defined as
\begin{equation}
\label{eq:16}
V_n(k,k^\prime)^2=\int\limits_0^{2\pi}\dfrac{d\theta}{2\pi}|\tilde{v}(\textbf{k}-\textbf{k}^\prime)|^2
\cos n\theta.
\end{equation}
Importantly, in Eq.~(\ref{eq:15}), we introduce a cut-off wave-vector $K_c=\pi/a_0$ (where $a_0$ is the lattice constant in the QW plane, which is actually the lattice constant of the CdTe buffer), corresponding to the edge of 2D Brillouin zone (cf.~Ref.~\cite{TAI3Dh}). This naturally limits the integration over quasimomentum~$k$.

Once the Green's function is known, one can calculate the spectral function $A(k,\varepsilon)$ and density-of-states $D(\varepsilon)$:
\begin{eqnarray}
\label{eq:17}
A(k,\varepsilon)=-\dfrac{1}{\pi}\textrm{Im}\left\{\textrm{Tr}\left(\hat{G}(k,\varepsilon+i0)\right)\right\},\notag\\
D(\varepsilon)=g_S\int\limits_0^{K_c}\dfrac{k dk}{2\pi}A(k,\varepsilon),~~~~~~~~~~
\end{eqnarray}
where the factor $g_S=2$ takes into account the contribution from the lower block in Eq.~(\ref{eq:1}). The density-of-states $D(\varepsilon)$ provides the most direct way to trace the evolution of the effective band-gap as a function of the impurity scattering, while $A(k,\varepsilon)$ represents the renormalization of the quasiparticle spectral properties at a given disorder strength.

As clear, Eqs.~(\ref{eq:11}) and (\ref{eq:15}) form a system of integral equations that defines the Green's function matrix $\hat{G}(k,\varepsilon)$. The self-consistent solution of such integral systems in the general case is a laborious task. However, for some particular cases of randomly distributed individual impurities, the solution of the problem can be greatly simplified. To proceed further, we assume $\tilde{v}(q)=u_0$,
which corresponds to the disorder formed by the short-range impurities:
\begin{equation}
\label{eq:18}
v(\textbf{r})=\int\limits_0^{\infty}\dfrac{qdq}{2\pi}\tilde{v}(q)\int\limits_0^{2\pi}\dfrac{d\theta}{2\pi}e^{i{q}{r}\cos\theta}=\dfrac{u_0}{2\pi}\dfrac{\delta\left(r\right)}{r}.
\end{equation}
In this case, $V_n(k,k^\prime)^2=u_0^2\delta_{n,0}$ in Eq.~(\ref{eq:16}), resulting in a diagonal form of the self-energy matrix in Eq.~(\ref{eq:15}):
\begin{widetext}
\begin{multline}
\label{eq:19}
\hat{\Sigma}(\varepsilon)=\begin{pmatrix}
\Sigma_{E1}(\varepsilon) & 0 & 0 & 0 \\[4pt]
0 & \Sigma_{H1}(\varepsilon) & 0 & 0 \\[4pt]
0 & 0 & \Sigma_{H2}(\varepsilon) & 0 \\[4pt]
0 & 0 & 0 & \Sigma_{E2}(\varepsilon)
\end{pmatrix}=
\dfrac{W^2}{4\pi}\int\limits_0^{K_c^2}dx
\begin{pmatrix}
G_{11}\left(\sqrt{x},\varepsilon\right) & 0 & 0 & 0 \\[4pt]
0 & G_{22}\left(\sqrt{x},\varepsilon\right) & 0 & 0 \\[4pt]
0 & 0 & G_{33}\left(\sqrt{x},\varepsilon\right) & 0 \\[4pt]
0 & 0 & 0 & G_{44}\left(\sqrt{x},\varepsilon\right)
\end{pmatrix},
\end{multline}
\end{widetext}
where we define the disorder strength as $W^2=n_{i}u_0^2$. We recall that $G_{ij}(k,\varepsilon)$ in Eq.~(\ref{eq:19}) are the component of the averaged Green's function in Eq.~(\ref{eq:11}), which in turns also depends on the self-energy. The self-energy independence from $k$ allows for an analytical calculation of the integrals in Eq.~(\ref{eq:19}).

Indeed, first of all, we note that the determinant of the matrix $\left[\varepsilon-\tilde{H}_{4\times4}(\sqrt{x})-\hat{\Sigma}(\varepsilon)\right]$ for the case of short-range impurities is represented as a polynomial of the fourth degree in $x$:
\begin{multline}
\label{eq:20}
\det\left|\varepsilon-\tilde{H}_{4\times4}(\sqrt{x})-\hat{\Sigma}(\varepsilon)\right|=A_4(\varepsilon)x^4+A_3(\varepsilon)x^3~\\
+A_2(\varepsilon)x^2+A_1(\varepsilon)x+A_0(\varepsilon),
\end{multline}
where $A_4(\varepsilon)$, $A_3(\varepsilon)$, $A_2(\varepsilon)$, $A_1(\varepsilon)$ and $A_0(\varepsilon)$ are complex functions, found by straightforward calculation of $4\times4$ symmetric matrix determinant.

Second, the diagonal components of the Green's function $G_{ii}\left(\sqrt{x},\varepsilon\right)$ ($i=1...4$) in Eq.~(\ref{eq:19}) are presented as
\begin{equation}
\label{eq:21}
G_{ii}\left(\sqrt{x},\varepsilon\right)=\dfrac{a_{i}(\varepsilon)x^3+b_{i}(\varepsilon)x^2+c_{i}(\varepsilon)x+d_{i}(\varepsilon)}
{\det\left|\varepsilon-\tilde{H}_{4\times4}(\sqrt{x})-\hat{\Sigma}(\varepsilon)\right|},
\end{equation}
where the terms $a_i$, ... $d_i$ are independent of $x$. This last equation  can be verified by the direct calculation of the inverse matrix $\left[\varepsilon-\tilde{H}_{4\times4}(\sqrt{x})-\hat{\Sigma}(\varepsilon)\right]^{-1}$. As the self-energy matrix has imaginary part, all the coefficients in Eq.~(\ref{eq:21}) are complex as well.

\begin{figure*}
\includegraphics [width=1.95\columnwidth, keepaspectratio] {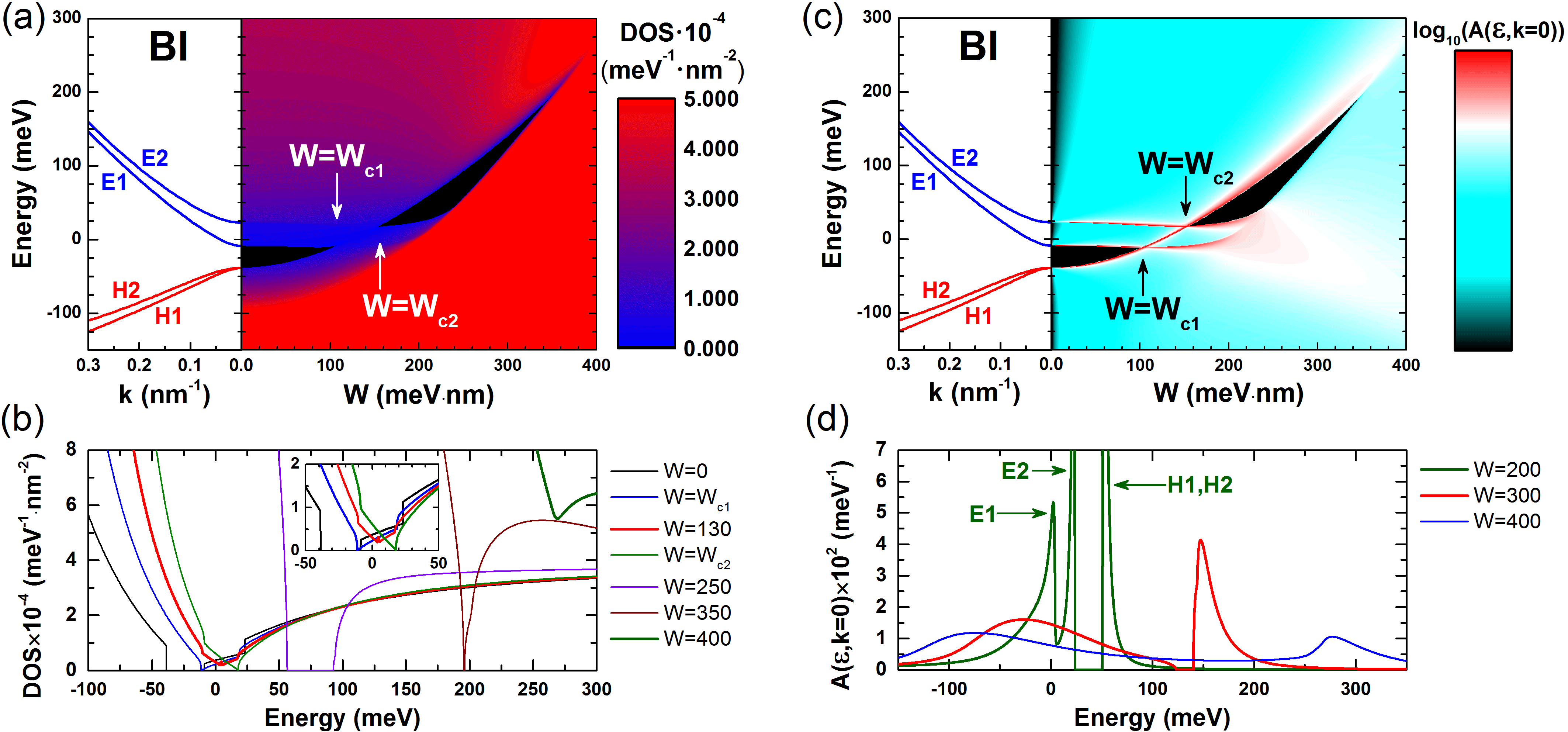} 
\caption{\label{Fig:3} (a) Band structure and color map of the DOS as a function of the disorder strength $W=\sqrt{{n}_{i}}u_0$ calculated for double HgTe~QW, which has a BI state in the clean limit (see the structure ``A'' in Fig.~\ref{Fig:1}). (b) The DOS as a function of energy at different values of $W$. The inset shows in more detail the DOS behavior at $W_{c1}{\leq}W{\leq}W_{c2}$. (c)~Band structure and color map of the spectral function at the $\Gamma$ point for the same double HgTe QW. (d) Energy dependence of $A(k=0,\varepsilon)$ at different disorder strength $W>W_{c2}$.}
\end{figure*}

In order to calculate the integrals in Eq.~(\ref{eq:19}), we have numerically found the roots $x_1$, $x_2$, $x_3$, $x_4$ of the polynomial needed for the following expansion:
\begin{multline}
\label{eq:22}
A_4x^4+A_3x^3+A_2x^2+A_1x+A_0~\\
=A_4(x-x_1)(x-x_2)(x-x_3)(x-x_4)
\end{multline}
with $A_0$, ... $A_4$ are independent of $x$. Although the values of $x_1$, $x_2$, $x_3$, $x_4$ can be found analytically by means of \emph{Ferrari's method}, the numerical procedures also allow for the calculations with any needed degree of accuracy. Once the roots are known, the integrals are calculated analytically (see Appendix). In this case, this self-consistent system of integral equations is transformed into the set of algebraic equations, which can be numerically solved by iteration procedure.

Finally, the spectral function $A(k,\varepsilon)$ and density-of-states $D(\varepsilon)$ for the case of the short-range disorder can be written as
\begin{eqnarray}
\label{eq:23}
A(k,\varepsilon)=-\dfrac{1}{\pi}\textrm{Im}\left\{ \sum\limits_{i=1}^{4}
G_{ii}\left(k,\varepsilon\right)\right\},\notag\\
D(\varepsilon)=-\dfrac{g_S}{W^2\pi}\textrm{Im}\left\{\textrm{Tr}\left(\hat{\Sigma}(\varepsilon)\right)\right\}.~
\end{eqnarray}
We remind that in order to take into account the contribution from the lower block in Eq.~(\ref{eq:1}), one should set the degeneracy factor of DOS as $g_S=2$.

We emphasize that in the case of short-range disorder, the self-energy matrix $\hat{\Sigma}(\varepsilon)$ retains its diagonal form also when taking into account the anisotropic part $H_{4\times4}^{(a)}(\mathbf{k})$. The latter, resulting  only in very small corrections, significantly complicates the self-consistent calculations, since it does not allow analytical calculation of the integrals in Eq.~(\ref{eq:13}). However, since the anisotropic part $H_{4\times4}^{(a)}(\mathbf{k})$ is due to crystal symmetry, it must always be taken into account in an identification of higher-order topological states~\cite{q23}.

\section{\label{Sec:RnD} Results and Discussion}
As known for ``conventional'' HgTe/CdHgTe QWs~\cite{q1,q2,q3,q4}, if the QW has a trivial band ordering in the clean limit, the presence of disorder leads to the band inversion and disorder-induced topological phase transition~\cite{q14,q15}. A similar behavior can also be expected for double HgTe QWs with a thick opaque barrier. In the latter case, the 2D system is described by two identical copies of Bernevig-Hughes-Zhang (BHZ) model~\cite{q1}, each of which corresponding to one of the HgTe QWs with parameters $S_0$ and $R_{1,2}^{(i)}$ set to zero for the inter-layer coupling (see Eq.~(\ref{eq:3})). In the presence of a tunnel-transparent barrier, that leads to a rich phase diagram shown in Fig.~\ref{Fig:1}, the BHZ-like blocks ``1'' and ``2'' of $H_{4\times4}(\mathbf{k})$ in Eq.~(\ref{eq:3}) cease to be identical and become coupled due to non-zero parameters $S_0$ and $R_{1,2}^{(i)}$. Nevertheless, even in this case, it is reasonable to expect a disorder-induced inversion both between both subband pairs (\emph{E}1,~\emph{H}1) and (\emph{E}2,~\emph{H}2).

Figure~\ref{Fig:3}(a) shows the evolution of DOS with the disorder strength $W=\sqrt{{n}_{i}}u_0$
for double HgTe~QW with BI state in the clean limit (the structure ``A'' in Fig.~\ref{Fig:2} and Tab.~\ref{Tab:1} with positive mass parameters $M_1$ and $M_2$). As clear, the band-gap in the DOS decreases with $W$ until it vanishes above a critical value $W_{c1}$, and then it is reopened again at $W>W_{c2}$ (see black regions in Fig.~\ref{Fig:3}(a) where the DOS vanishes). In the range $W_{c1}{\leq}W{\leq}W_{c2}$, the double QW remains gapless, which is well illustrated by the non-zero DOS in Fig.~\ref{Fig:3}(b). With a further increase in disorder strength, the newly opened gap closes again at $W\simeq350$~meV$\cdot$nm. Further, we demonstrate that such complex behavior of the DOS is directly attributed to the disorder-induced topological phase transitions in double HgTe QWs.

After theoretical explanation of disorder-induced phase transition in single HgTe QWs in terms of the effective renormalization of topological mass and chemical potential~\cite{q15}, one has a natural desire to describe such disorder-induced transitions within the framework of the concept of \emph{quasiparticles}. Assuming that the Green's function for a disordered system $\hat{G}_{2D}(\mathbf{k},\varepsilon)$ is known, the quasiparticle Hamiltonian is defined as follows
\begin{equation}
\label{eq:B1}
\hat{G}_{\mathrm{2D}}(\mathbf{k},\varepsilon)=
\left[\varepsilon-\mathcal{H}_{qp}(\mathbf{k},\varepsilon)\right]^{-1},
\end{equation}
with
\begin{equation}
\label{eq:B2}
\mathcal{H}_{qp}(\mathbf{k},\varepsilon)=
H_{\mathrm{2D}}(\mathbf{k})+\hat{\Sigma}_{\mathrm{2D}}(\mathbf{k},\varepsilon),
\end{equation}
where $H_{2D}(\mathbf{k})$ is the Hamiltonian of the clean system, while the self-energy $\hat{\Sigma}_{\mathrm{2D}}(\mathbf{k},\varepsilon)$ encodes the interaction with the disorder potential. Quasiparticle Hamiltonian for the case of double HgTe QWs in the presence of short-range disorder can be presented in the form
\begin{equation}
\label{eq:B3}
\mathcal{H}_{qp}(\mathbf{k},\varepsilon)=\begin{pmatrix}
H_{4\times4}(\mathbf{k}) & 0 \\ 0 & H_{4\times4}^{*}(-\mathbf{k})\end{pmatrix}
+\begin{pmatrix}
\hat{\Sigma}(\varepsilon) & 0 \\ 0 & \hat{\Sigma}(\varepsilon)\end{pmatrix},
\end{equation}
where the blocks $H_{4\times4}(\mathbf{k})$ and $\hat{\Sigma}(\varepsilon)$ are given by Eqs.~(\ref{eq:2}) and (\ref{eq:19}), respectively.

In the most general case, $\mathcal{H}_{qp}(\mathbf{k},\varepsilon)$ is \emph{non-Hermitian} -- the real part of its eigenvalues characterizes the ``renormalized'' band structure, while the imaginary part describes the quasiparticle decay. The latter can be illustrated by the form of the spectral function $A(\mathbf{k},\varepsilon)$, which is written in terms of the eigenvalues $\mathcal{E}_{qp}^{(n)}(\mathbf{k},\varepsilon)$ of $\mathcal{H}_{qp}(\mathbf{k},\varepsilon)$ as follows:
\begin{equation}
\label{eq:B4}
A(\mathbf{k},\varepsilon)=-\dfrac{1}{\pi}\sum\limits_{n}
\dfrac{\textrm{Im}\mathcal{E}_{qp}^{(n)}}
{\left(\varepsilon-\textrm{Re}\mathcal{E}_{qp}^{(n)}\right)^2
+\left(\textrm{Im}\mathcal{E}_{qp}^{(n)}\right)^2},
\end{equation}
where $n$ is multi-index that labels quasiparticle subbands $E1_{\pm}$, $H1_{\pm}$, $H2_{\pm}$, $E2_{\pm}$. Here, the signs ``+'' and ``-'' refer to the states of the upper and lower blocks of $\mathcal{H}_{qp}(\mathbf{k},\varepsilon)$ in Eq.~(\ref{eq:B3}) respectively. In the clean limit (when $W=0$), all $\mathcal{E}_{qp}^{(n)}(\mathbf{k},\varepsilon)$ vanish, and the spectral function is represented by sum of $\delta$-functions centered at the the eigenvalue energies of $H_{4\times4}(\mathbf{k})$ in Eq.~(\ref{eq:2}). As $W$ increases, the spectral function $A(\mathbf{k},\varepsilon)$ widens -- its maxima, determined from the condition $\varepsilon=\textrm{Re}\mathcal{E}_{qp}^{(n)}(\mathbf{k},\varepsilon)$, still represent the quasiparticle subband dispersion~\cite{TAI3Dh}, while its broadening in the vicinity of the maxima values define the quasiparticle lifetime.

Figure~\ref{Fig:3}(c) provides a disorder-induced evolution of the spectral function at the $\Gamma$ point, which perfectly reproduces two band-gap regions arising in the DOS. As clear, the first band-gap closing at $W=W_{c1}$ is indeed attributed to the mutual inversion of \emph{E}1 and \emph{H}1 quasiparticle subbands, while the band-gap opening at $W=W_{c2}$ is related to the band inversion but in the second pair of \emph{E}2 and \emph{H}2 quasiparticle subbands. For a better understanding of the origin of the second band-gap closing at $W>W_{c2}$, Figure~\ref{Fig:3}(d) shows an energy dependence of the spectral function at the $\Gamma$ point at different disorder strength. As seen, if the disorder strength only slightly exceeds the critical value $W_{c2}$, the spectral function shows pronounced peaks associated with quasiparticles from different subbands. Here, we underline the strongly asymmetrical non-Lorentzian shape of the peaks in $A(k=0,\varepsilon)$. As $W$ increases, the peaks corresponding to the \emph{E}1 and \emph{E}2 subbands at $W>W_{c2}$ broaden and merge into one, which shifts towards lower energies, while the peak defining the quasiparticles in the \emph{H}1 subband shifts towards higher energies. Finally, with a further growth of $W$, the broadening of both peaks, corresponding to the electron-like and hole-like states, increases by so much that it leads to the band-gap closing at $W>350$~meV$\cdot$nm. This signals that no coherent quasiparticles can be defined in this case, and incoherent processes associated with their decay are relevant at a high disorder strength.

\begin{figure*}
\includegraphics [width=2.05\columnwidth, keepaspectratio] {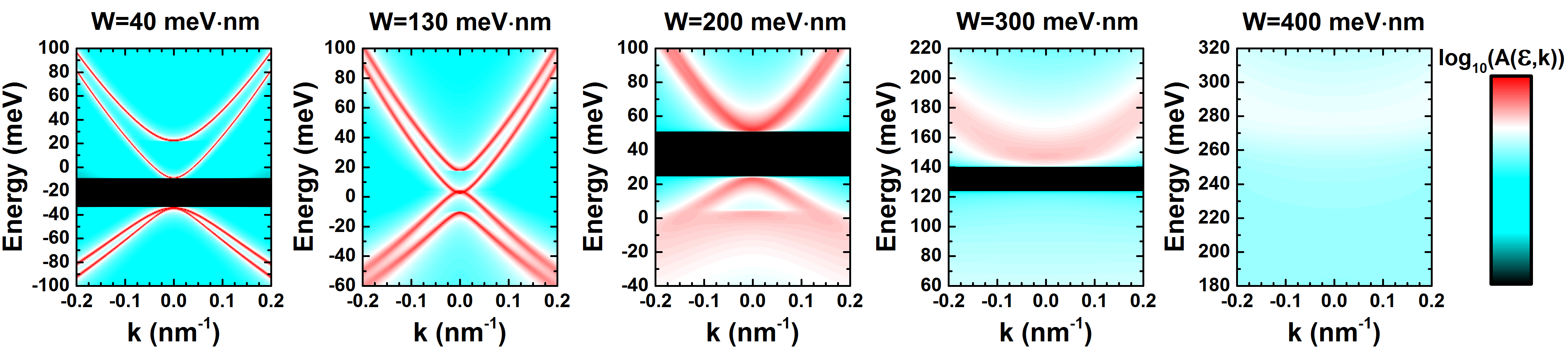} 
\caption{\label{Fig:4} Colormap of spectral function at different disorder strength for double HgTe~QW, which has a BI state in the clean limit (see the structure ``A'' in Fig.~\ref{Fig:1}).}
\end{figure*}

Figure~\ref{Fig:4} shows the evolution of the spectral function $A(\mathbf{k},\varepsilon)$ with increasing  disorder strength. As clear, a gapless state at $W_{c1}{<}W{<}W_{c2}$ with the band ordering \emph{E}2-\emph{H}1-\emph{H}2-\emph{E}1 can be directly attributed to the BG state in the clean limit~\cite{q22}. Moreover, the transformation of the quasiparticle spectrum with $W$ indeed resembles the changes in the band structure of double HgTe QW with $d$ (see Fig.~\ref{Fig:2}). Thus, starting from a band insulator in the clean limit, the 2D system, under influence of disorder, successively passes into the BG state, and then into the insulator with double-band inversion. However, in order to prove that $W=W_{c1}$ and $W=W_{c2}$ indeed represent the topological phase transitions, one should demonstrate a change in the $\mathbb{Z}_2$ invariant, as it was done in Section~\ref{Sec:Diagram} for the clean limit.

Until now, the calculation of various topological invariants in 2D systems with interactions was carried out on the basis of the Green's functions $\hat{G}_{\mathrm{2D}}(\mathbf{k},\varepsilon)$~\cite{q37,q38,q39,q40}. Alternatively, the calculation of $\mathbb{Z}_2$ invariants (as well as other invariants) is much more convenient and visually performed by means of $\mathcal{H}_{qp}(\mathbf{k},\varepsilon)$ based on methods developed for the non-Hermitian case in recent years~\cite{newR1,newR2,newR3,newR4,newR5,newR6}. Similar to the Hermitian case, topological invariants for any quasiparticle band $\mathcal{E}_{qp}^{(n)}(\mathbf{k},\varepsilon)$ in 2D system, can be constructed from the associated eigenstates of $\mathcal{H}_{qp}(\mathbf{k},\varepsilon)$~\cite{newR1,newR2,newR3}. A block-diagonal form of $\mathcal{H}_{qp}(\mathbf{k},\varepsilon)$ in Eq.~(\ref{eq:B3}) reduces the calculation of the $\mathbb{Z}_2$ invariant to the calculation of
\begin{equation}
\label{eq:B5}
\mathbb{Z}_2(\varepsilon)=\mathrm{mod}\left[\dfrac{C_{\uparrow}(\varepsilon)-C_{\downarrow}(\varepsilon)}{2},2\right],
\end{equation}
where $C_{\uparrow}(\varepsilon)$ and $C_{\downarrow}(\varepsilon)$ are the Chern numbers for the upper and lower blocks, respectively.

Importantly, since $\mathcal{H}_{qp}(\mathbf{k},\varepsilon)$ is non-Hermitian, their left and right eigenstates are generally unrelated and satisfy the following eigenvalue equations:
\begin{eqnarray}
\label{eq:B6}
\mathcal{H}_{qp}|\Psi_n^{R}(\mathbf{k},\varepsilon)\rangle=
\mathcal{E}_{qp}^{(n)}|\Psi_n^{R}(\mathbf{k},\varepsilon)\rangle,~~~~~\notag\\
\left(\mathcal{H}_{qp}\right)^{\dag}(\mathbf{k},\varepsilon)|\Psi_n^{L}(\mathbf{k},\varepsilon)\rangle=
{\mathcal{E}_{qp}^{(n)}}^{*}|\Psi_n^{L}(\mathbf{k},\varepsilon)\rangle.
\end{eqnarray}
Thus, for any complex subband $\mathcal{E}_{qp}^{(n)}$, one can construct four different gauge invariant Berry curvatures~\cite{newR1}:
\begin{equation}
\label{eq:B7}
B_{n,ij}^{\alpha\beta}(\mathbf{k},\varepsilon)=i\langle\partial_{i}\Psi_n^{\alpha}(\mathbf{k},\varepsilon)|\partial_{j}\Psi_n^{\beta}(\mathbf{k},\varepsilon)\rangle
\end{equation}
with the normalization condition $\langle\Psi_n^{\alpha}\rangle|\Psi_n^{\beta}\rangle=1$, where $i,j\in\left\{k_x,k_y\right\}$ and $\alpha,\beta\in\left\{L,R\right\}$. Note that although these  ``left-left'', ``left-right'', ``right-left'' and ``right-right'' Berry curvatures are locally different quantities, their integrals all yield the same Chern number for a given subband $\mathcal{E}_{qp}^{(n)}$~\cite{{newR1}}:
\begin{equation}
\label{eq:B8}
C_{n}(\varepsilon)=\dfrac{i}{2\pi}{\int}\epsilon_{ij}B_{n,ij}^{\alpha\beta}(\mathbf{k},\varepsilon)
{d^2\textbf{k}},
\end{equation}
where $\epsilon_{ij}=-\epsilon_{ji}$ denotes the Levi-Civita symbol in 2D and the summation over $i$ and $j$ is implied. By summing up the Chern numbers of the ``valence'' quasiparticle subbands, identified through the spectral function $A(\mathbf{k},\varepsilon)$ (see Fig.~\ref{Fig:4}), one can calculate $C_{\uparrow}(\varepsilon)$ and $C_{\downarrow}(\varepsilon)$ and, accordingly, the $\mathbb{Z}_2$ invariant. The latter defines the regions on the plane ($\varepsilon$,$W$), where the $\mathbb{Z}_2$ invariant is equal to $0$ or $1$.

To the best of our knowledge, no one has yet calculated any topological invariants in 2D systems with disorder on the basis of the quasiparticle Hamiltonian. Therefore, Supplemental Materials~\cite{SM} demonstrate an application of the general concept above for the case of single HgTe QWs described within the \emph{continuous} BHZ model. The latter allows one to analytically calculate the $\mathbb{Z}_2$ invariant and the conductance inside the band-gap in the presence of short-range disorder. In what follows, we focus on the more complex case of double HgTe QWs with the short-range disorder.

As was shown by Kawabata~\emph{et~al.}~\cite{q41}, if non-Hermitian Hamiltonian respects both time-reversal and inversion symmetries, the $\mathbb{Z}_2$ topological invariant can be defined just like in the Hermitian case (cf. Eq.~(\ref{eq:4ad4})). This reduces the calculation of the $\mathbb{Z}_2$ invariant for $\mathcal{H}_{qp}(\mathbf{k},\varepsilon)$ in Eq.~(\ref{eq:B3}) to a parity analysis of the quasiparticle subbands at the $\Gamma$ point, just as was done in Sec.~\ref{Sec:Diagram}. Importantly, the diagonal form of the self-energy matrix $\hat{\Sigma}_{\mathrm{2D}}(\varepsilon)$ in Eq.~(\ref{eq:B3}) and its independence from $\textbf{k}$ mean that the presence of short-range disorder does not affect the parity of the quasiparticle subbands. In other words, the quasiparticle subband parities coincide with those in the clean limit (cf.  Eq.~(\ref{eq:4ad3})). By means of the arguments similar to the ones used in Sec.~\ref{Sec:Diagram}, the calculation of the $\mathbb{Z}_2$ invariant on the basis of the parity analysis is reduced to
\begin{equation}
\label{eq:B9}
-(-1)^{\mathbb{Z}_2(\varepsilon)}=\prod\limits_{n:~\mathrm{valence~of~}{\mathcal{H}_{qp}(\mathbf{k},\varepsilon)}}
\pi_{n}\left(0\right),
\end{equation}
where $\pi_{n}\left(0\right)\in\{\pm{1}\}$ is the parity eigenvalue of the $n$-th Kramers pair of quasiparticle subbands at the $\Gamma$ point defined by Eq.~(\ref{eq:4ad3}). Note that the left and right eigenstate corresponding to the same eigenvalue $\mathcal{E}_{qp}^{(n)}$ has the same parity, therefore both of them can be used for calculating $\pi_{n}\left(0\right)$ in Eq.~(\ref{eq:B9}). Thus, the $\mathbb{Z}_2$ topological invariant indeed changes its value at $W=W_{c1}$ and $W=W_{c2}$, that represents topological phase transitions into BG state and double-inverted insulator, respectively. Interestly, the band-gap closing at $W>W_{c2}$ formally corresponds to $\mathbb{Z}_2(\varepsilon)=0$ in accordance with Eq.~(\ref{eq:B9}). However, a feature of such band-gap closing, as noted earlier, is not the subband inversion, but the overlap of the spectral functions from \emph{E}1 and \emph{H}1 quasiparticle subbands (the latter will become clear later). This indicates that the imaginary parts of $\mathcal{E}_{qp}^{(n)}$ significantly exceed the real parts, which determine the quasiparticle subband dispersions. Therefore, this type of the band-gap closing is apparently associated with a change in the \emph{vorticity} -- a topological invariant defined for any pair of the bands as the winding number of their energies $\mathcal{E}_{qp}^{(n)}$ in the complex-energy plane~\cite{newR1}. A detailed clarification of this point is beyond the scope of the current work and will be the subject of future research.

Let us now demonstrate that a double-inverted insulator state at $W>W_{c2}$ is actually a HOTI with the corner states inside the band-gap. To prove the existence of the corner states, on must take into account the anisotropic part ${H}_{4\times4}^{(a)}(\mathbf{k})$ given by Eq.~(\ref{eq:4ad2}) and associated with the cubic crystal symmetry. Then, by applying open boundary conditions in quasiparticle Hamiltonian $\mathcal{H}_{qp}(\mathbf{k},\varepsilon)$, one can derive an effective 1D Hamiltonian $\mathcal{H}_{qp}^{\mathrm{edge}}(\mathbf{k},\varepsilon)$ for the quasiparticle edge states \emph{inside the band-gap}, where the quasimomentum $\mathbf{k}$ is oriented along the edge. Importantly, since $A(\mathbf{k},\varepsilon)$ vanishes in the band-gap regions (see Figs~\ref{Fig:3}(c) and \ref{Fig:4}), $\textrm{Im}\mathcal{E}_{qp}^{(n)}(\mathbf{k},\varepsilon)=0$, and thus $\hat{\Sigma}(\varepsilon)$ has only \emph{real components} inside the band-gap. Therefore, the 1D edge Hamiltonian $\mathcal{H}_{qp}^{\mathrm{edge}}(\mathbf{k},\varepsilon)$ is \emph{Hermitian}. The latter means that the edge quasiparticles \emph{do not decay} and their spectral function is represented as
\begin{multline}
\label{eq:B10}
A^{\mathrm{edge}}(\mathbf{k},\varepsilon)=-\dfrac{1}{\pi}\sum_{m}\textrm{Im}\dfrac{1}{\varepsilon-\mathcal{E}_{m}^{\mathrm{edge}}(\mathbf{k},\varepsilon)+i0^{+}}\\
=\sum_{m}\delta\left\{\varepsilon-\mathcal{E}_{m}^{\mathrm{edge}}(\mathbf{k},\varepsilon)\right\},
\end{multline}
where $m$ labels the eigenvalues $\mathcal{E}_{m}^{\mathrm{edge}}(\mathbf{k},\varepsilon)$ of the 1D edge Hamiltonian $\mathcal{H}_{qp}^{\mathrm{edge}}(\mathbf{k},\varepsilon)$.

Strictly speaking, the only difference between $\mathcal{H}_{qp}^{\mathrm{edge}}(\mathbf{k},\varepsilon)$ and the edge Hamiltonian in the clean limit~\cite{q23} is the renormalization of the structure parameters:
\begin{eqnarray}
\label{eq:B11}
C_1{\rightarrow}C_1+\dfrac{1}{2}\left[\textrm{Re}\Sigma_{E1}(\varepsilon)+\textrm{Re}\Sigma_{H1}(\varepsilon)\right],~\notag\\
M_1{\rightarrow}M_1+\dfrac{1}{2}\left[\textrm{Re}\Sigma_{E1}(\varepsilon)-\textrm{Re}\Sigma_{H1}(\varepsilon)\right],~\notag\\
C_2{\rightarrow}C_2+\dfrac{1}{2}\left[\textrm{Re}\Sigma_{E2}(\varepsilon)+\textrm{Re}\Sigma_{H2}(\varepsilon)\right],~\notag\\
M_2{\rightarrow}M_2+\dfrac{1}{2}\left[\textrm{Re}\Sigma_{E2}(\varepsilon)-\textrm{Re}\Sigma_{H2}(\varepsilon)\right],~
\end{eqnarray}
where the component of the self-energy matrix is given by Eq.~(\ref{eq:19}). Note again that since $\textrm{Im}\hat{\Sigma}(\varepsilon)=0$ inside the band-gap, the self-energy matrix components in the quasiparticle edge Hamiltonian are pure real. Thus, with the replacement expressed by Eq.~(\ref{eq:B11}), the problem of corner states for $\mathcal{H}_{qp}^{\mathrm{edge}}(\mathbf{k},\varepsilon)$ in the presence of the disorder is reduced to the problem already solved in the clean limit~\cite{q23}.

\begin{figure*}
\includegraphics [width=1.99\columnwidth, keepaspectratio] {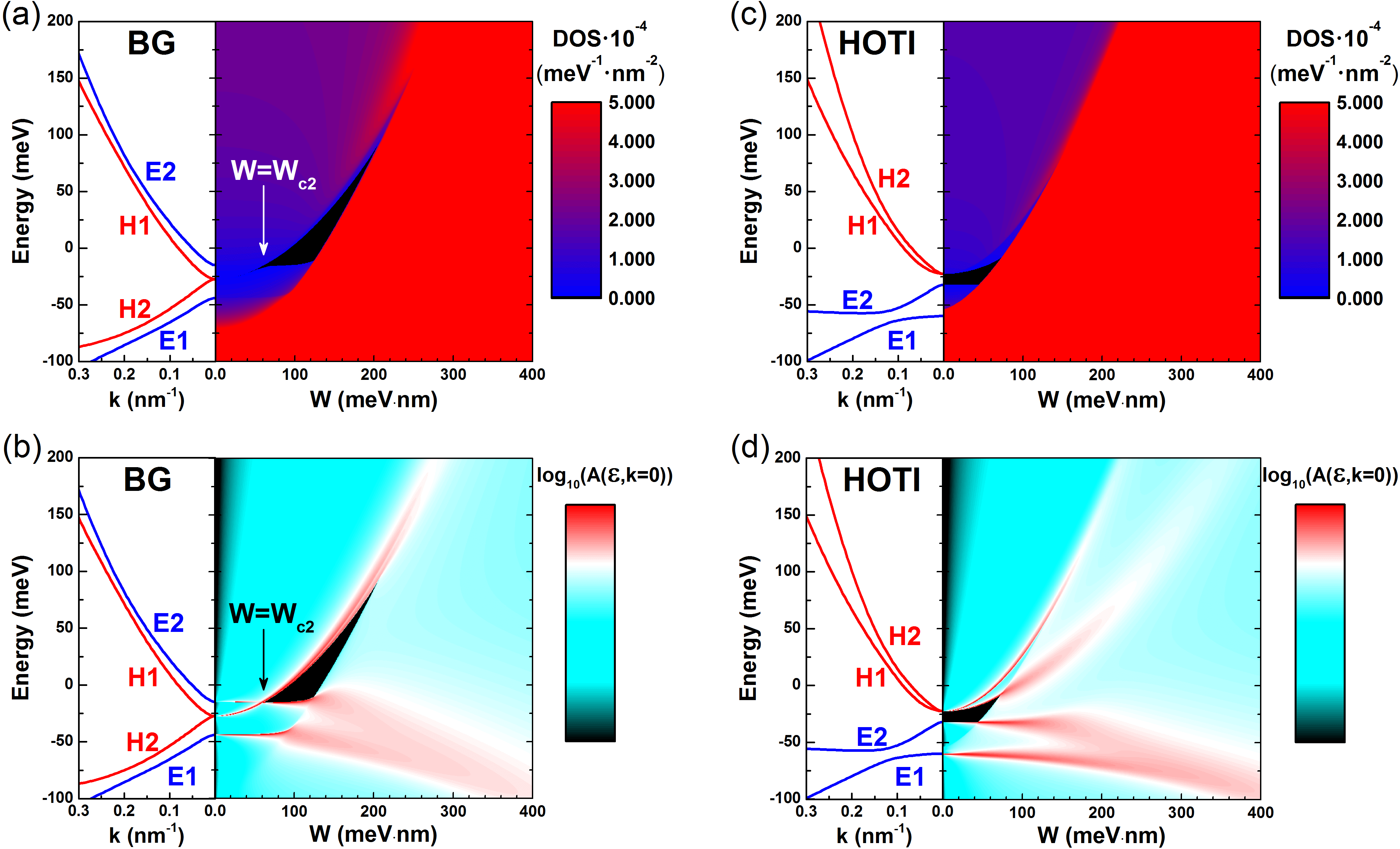} 
\caption{\label{Fig:5} (a,b) Band structure, DOS and spectral function  at the $\Gamma$ point $A(k=0,\varepsilon)$ as a function of the disorder strength $W=\sqrt{{n}_{i}}u_0$ for double HgTe QW hosting BG state at $W=0$ ($M_1<0$ and $M_2>0$, the structure ``C'' in Fig.~\ref{Fig:1}). (c,d) The same as in (a) and (b) panels but for double HgTe QW hosting HOTI state in the absence of disorder ($M_1<0$ and $M_2<0$, the structure ``E'' in Fig.~\ref{Fig:1}).}
\end{figure*}

In particular, taking into account the anisotropic part ${H}_{4\times4}^{(a)}(\mathbf{k})$ in the bulk quasiparticle Hamiltonian $\mathcal{H}_{qp}(\mathbf{k},\varepsilon)$ leads to a specific mass-kink of the \emph{gapped} Dirac edge states as function of the edge orientation. Such a mass-kink results in the appearance of a localized 0D state at the corner formed by the meeting edges, whose bisectors coincide with the mirror symmetry planes of (001) double HgTe QWs~\cite{q23}. Interestly, the Hermitian property of $\mathcal{H}_{qp}^{\mathrm{edge}}(\mathbf{k},\varepsilon)$ also implies that the quasiparticle corner states do not decay -- their spectral function representing the energies on the ($\varepsilon$,$W$) diagram is written as:
\begin{equation}
\label{eq:B12}
A^{\mathrm{corner}}(\varepsilon)=\sum_{\pm}\delta\left\{\varepsilon-\mathcal{E}_{\pm}^{\mathrm{corner}}(\varepsilon)\right\},
\end{equation}
where ``$\pm$'' takes into account the double degeneracy of the corner state energy due to the time reversal symmetry: $\mathcal{E}_{+}^{\mathrm{corner}}(\varepsilon)=\mathcal{E}_{-}^{\mathrm{corner}}(\varepsilon)$.
The latter can be directly calculated on the basis of the results obtained for the clean limit~\cite{q23} by taking into account Eq.~(\ref{eq:B11}).

Let us now briefly consider disorder-induced phase transitions in double HgTe QWs with single- and double-band-inversion at $W=0$ (see the diagram in Fig.~\ref{Fig:1}). Figure~\ref{Fig:5} provides the evolution of the DOS and the spectral function at the $\Gamma$ point with the disorder strength for the structures ``C'' and ``E'', hosting BG and HOTI states in the clean limit, respectively. As seen from Fig.~\ref{Fig:5}(a), the double HgTe QW with BG state maintains zero band-gap even in the presence of weak disorder. When the disorder strength reaches its critical value $W_{c2}$, a band-gap opens in the DOS. As clear from Fig.~\ref{Fig:5}(b), this moment corresponds to the band inversion between \emph{E}2 and \emph{H}2 subbands. Such behavior of the spectral function unambiguously indicates that the 2D system at $W<W_{c2}$ is in the same topological state as at $W=0$, which is uniquely identified as the BG state. When $W$ exceeds its critical value $W_{c2}$, the structure ``C'' becomes a HOTI (as it has been proved above), whose bulk band-gap vanishes as the disorder strength increases further (cf. Fig.~\ref{Fig:3}). A similar band-gap evolution is also seen for the structure ``E'' with a HOTI at $W=0$ as shown in Fig.~\ref{Fig:5}(c).

As clear from Fig.~\ref{Fig:5}(d), such band-gap behavior is caused primarily by smearing of the edges of the \emph{E}1 and \emph{H}1 bands at the $\Gamma$ point that leads to the DOS overlapping between conduction and valence subbands.

Finally, let us discuss the application of the obtained results to the real samples~\cite{Tr1,Tr2,Tr3,Tr4,Tr5,Tr6,Tr7,LL1,LL1b,LL2,LL3} with double HgTe QWs. As clear from Fig.~\ref{Fig:2}, the effective 2D low-energy model used for the calculation of DOS and spectral function can actually be applied to the states with the small values of $\mathbf{k}$. In this sense, the low-energy model used for double HgTe QWs is analogous to the BHZ model~\cite{q1} proposed earlier for the low-energy states of single HgTe QWs in the vicinity of the $\Gamma$ point of the Brillouin zone. At the same time, the calculation of the self-energy matrix within SCBA for the case of short-range scatterers requires the summation of all states in the entire Brillouin zone of \emph{real double HgTe QWs}. The latter also involves a large range of quasimomentum values, far beyond the applicability of the effective 2D low-energy model.

We emphasize that such an internal contradiction is also inherent in all previous studies performed on the basis of the BHZ model including its tight-binding version~\cite{q14,q15,q16,q17,q17b}. Indeed, regularization of the effective 2D low-energy model on a square lattice just leads to a change in the description at large quasimomentum by introducing a new Brillouin zone determined by the \emph{artificial} lattice, which differs significantly from the real lattice of HgTe. Moreover, in order for the continuous model and its tight-binding version to give the same results at small quasimomentum, the artificial lattice constant must differ by an order from the lattice constant of HgTe~\cite{q14,q15,q16,q17}. Therefore, the tight-binding version of the low-energy model also cannot be applied to describe the short-range disorder effects in real HgTe-based QW structures. This perhaps explains why disorder-induced phase transitions theoretically studied in the previous works have never been observed in solid state systems. Additionally, it is physically difficult to fine-tune the disorder in real semiconductor devices.

Fortunately, there are several non-solid-state systems implemented in cold atoms, photonic crystals and electrical circuits, whose entire Brillouin zone is fully described by a tight-binding version of the BHZ model. Recent experimental observation of TAI state in these systems~\cite{q32,q33,q34,q35} evidences that the results of current work can also be of interest for experimental verification in the ``analogues'' of double HgTe QWs also made on the basis of cold atoms~\cite{q32}, photonic crystals~\cite{q33} and electric circuits~\cite{q34,q35}.

\section{\label{Sec:Sum} Summary}
We have investigated the effect of disorder induced by short-range impurities on the band-gap of double HgTe QWs hosting trivial, single-band-inverted and double-band-inverted states in the clean limit. By using the SCBA and a four-band 2D low-energy Hamiltonian, we directly calculate the DOS and spectral function describing the quasiparticles at the $\Gamma$ point of the Brillouin zone.
By following the DOS and spectral function evolution when increasing the disorder strength, we unambiguously demonstrated multiple topological phase transitions caused by the mutual inversion of both first and second electron-like and hole-like subbands. We have shown that starting from a band insulator in the clean limit, the double HgTe QW under the disorder influence, successively passes into a BG state, and then into a HOTI state with a double-band inversion. We have found that all disorder-induced transitions are completely characterized by introducing a non-Hermitian quasiparticle Hamiltonian encoding the band structure renormalization and quasiparticle decay. Experimental observations of the disorder-induced phase transitions in non-solid-state systems based on cold atoms, photonic crystals and electric circuits evidence that the results of current work can also be of interest for experimental verification in the ``analogues'' of double HgTe QWs made on the basis of the mentioned systems.


\begin{acknowledgments}
This work was supported by the Terahertz Occitanie Platform, by the CNRS through IRP ``TeraMIR'' by the French Agence Nationale pour la Recherche (``Colector'' project) and the Russian Science Foundation (Grant \# 22-22-00382).
\end{acknowledgments}

\appendix*
\section{\label{sec:App} Calculation of integrals in Eq.~(\ref{eq:19})}
In order to calculate the integrals in Eq.~(\ref{eq:19}), one should find four complex roots $x_1(\varepsilon)$, $x_2(\varepsilon)$, $x_3(\varepsilon)$ and $x_4(\varepsilon)$ of the polynomial in Eq.~(\ref{eq:20}):
\begin{multline*}
\det\left|\varepsilon-\tilde{H}_{4\times4}(\sqrt{x})-\hat{\Sigma}(\varepsilon)\right|=A_4(\varepsilon)x^4+A_3(\varepsilon)x^3~\\
+A_2(\varepsilon)x^2+A_1(\varepsilon)x+A_0(\varepsilon).
\end{multline*}
The latter is derived by straightforward calculation of $4\times4$ symmetric matrix determinant:
\begin{widetext}
\begin{multline}
\label{eq:A1}
\det\left|\hat{A}\right|=a_{12}^2a_{34}^2-a_{33}a_{44}a_{12}^2+2a_{44}a_{12}a_{13}a_{23}-2a_{12}a_{13}a_{24}a_{34}-2a_{12}a_{14}a_{23}a_{34}+2a_{33}a_{12}a_{14}a_{24}-a_{22}a_{33}a_{14}^2~~~~~~~~\\
+a_{13}^2a_{24}^2-a_{22}a_{44}a_{13}^2-2a_{13}a_{14}a_{23}a_{24}+2a_{22}a_{13}a_{14}a_{34}+a_{14}^2a_{23}^2
-a_{11}a_{44}a_{23}^2+2a_{11}a_{23}a_{24}a_{34}-a_{11}a_{33}a_{24}^2~\\
-a_{11}a_{22}a_{34}^2+a_{11}a_{22}a_{33}a_{44}.~~~~
\end{multline}
Once the roots are known, the \emph{complex} integrals in Eq.~(\ref{eq:19}) are calculated as
\begin{multline}
\label{eq:Bsm12}
\int\dfrac{ax^3+bx^2+cx+d}{A_4(x-x_1)(x-x_2)(x-x_3)(x-x_4)}dx=\dfrac{ax_1^3+bx_1^2+cx_1+d}{A_4(x_1-x_2)(x_1-x_3)(x_1-x_4)}\ln\left(x-x_1\right)~~~~~~~~~~~~~~~~~~~~~~~~~\\
+\dfrac{ax_2^3+bx_2^2+cx_2+d}{A_4(x_2-x_1)(x_2-x_3)(x_2-x_4)}\ln\left(x-x_2\right)
+\dfrac{ax_3^3+bx_3^2+cx_3+d}{A_4(x_3-x_1)(x_3-x_2)(x_3-x_4)}\ln\left(x-x_3\right)~~~~~~\\
+\dfrac{ax_4^3+bx_4^2+cx_4+d}{A_4(x_4-x_1)(x_4-x_2)(x_4-x_3)}\ln\left(x-x_4\right).
\end{multline}
\end{widetext}

%

\end{document}